\documentclass[pre,twocolumn,superscriptaddress,nofootinbib]{revtex4-1}

\newcommand\ba{\begin{eqnarray}}
\newcommand\ea{\end{eqnarray}}
\newcommand\be{\begin{equation}}
\newcommand\ee{\end{equation}}

\newcommand{\ket}[1]{|#1\rangle}

\def\non{\nonumber}

\usepackage{amsfonts}
\usepackage{graphicx}
\usepackage{bm}
\usepackage{amssymb}
\usepackage{color}
\usepackage{amsmath}
\usepackage{amstext}
\usepackage{latexsym}
\usepackage[usenames,dvipsnames]{xcolor}
\usepackage[colorlinks=true,citecolor=Cerulean,linkcolor=RubineRed,urlcolor=Cerulean]{hyperref}

\def\non{\nonumber}

\begin{document}
\title{Bath Engineering Enhanced Quantum Critical Engines}

\author{Revathy B. S}
\affiliation{Department of Physics,  Indian Institute of Technology Palakkad,  Palakkad, 678557,  India}
\author{Victor Mukherjee}
\affiliation{Department of Physical Sciences, IISER Berhampur, Berhampur 760010, India}
\author{Uma Divakaran}
\affiliation{Department of Physics,  Indian Institute of Technology Palakkad,  Palakkad, 678557,  India}

\date{\today}

\begin{abstract}
Driving a quantum system across quantum critical points leads to non-adiabatic excitations in the system. This in turn may adversely affect the functioning of a quantum machine which uses a quantum critical substance as its working medium. Here we propose {a {\it bath-engineered quantum engine}} (BEQE), in which we use the Kibble--Zurek mechanism {{and critical scaling laws}} to formulate a protocol for enhancing the performance of finite-time quantum engines operating close to quantum phase transitions. In the case of free fermionic systems, BEQE enables finite-time engines to outperform engines operating in the presence of shortcuts to adiabaticity, and even infinite-time engines under suitable conditions, thus showing the remarkable advantages offered by this technique. Open questions remain regarding the use of BEQE based on non-integrable~models.
\end{abstract}


\maketitle

\section{Introduction}

The field of quantum thermodynamics aims to form a coherent understanding of the thermodynamics of quantum systems \cite{Alicki2018, doi:10.1080/00107514.2016.1201896, gemmer2009quantum, e15062100, binder2018thermodynamics, bhattacharjee21quantum, myers22quantum}. In classical thermodynamics, one can then use this knowledge to understand the limitations on the performance of quantum machines. In this respect, quantum control can play a significant role in enabling us to go beyond these limitations and develop high-performing quantum machines 
\cite{erdman22identifying, PhysRevResearch.4.L012029}. This can be especially significant in the case of finite-time quantum machines \cite{PhysRevE.76.031105, Mukherjee_2021}, as non-adiabatic excitations can be detrimental to the performances of such machines, thus necessitating the application of controls in order to boost their outputs \cite{GELBWASERKLIMOVSKY2015329, mukherjee17enhanced}. 

Control techniques such as shortcuts to adiabaticity (STA) have been shown to be highly successful in enhancing the output of finite-time quantum engines \cite{delcampo14, PhysRevResearch.2.023145, Deng13, Beau16, chenu18thermodynamics, sels17minimizing}. However, the application of STA can be highly non-trivial in many-body quantum engines, owing to the diverging dimensions of the associated Hilbert spaces. This can be especially challenging in quantum engines operating close to quantum critical points, where the diverging length and time scales can demand STA protocols involving long-range interactions \cite{campo12assisted, kolodrubetz17geometry}. The above challenges motivated us to search for control protocols beyond STA for application in quantum engines operating close to quantum phase transitions. 

In this work we propose a control protocol aimed at enhancing the efficiency as well as the output work of quantum engines based on free fermionic working mediums {{(WMs)}} operating close to quantum critical points \cite{sachdev_2011}. 
Quantum phase transitions have proven to be beneficial for quantum heat engines \cite{campisi2016power, Yang19, PhysRevE.96.022143, piccitto2022ising, Fogarty_2020}. The universality in quantum critical machines arising from the Kibble--Zurek mechanism (KZM) has already been studied in~\cite{revathy20universal}.
Here, we construct a quantum heat engine using a working medium that undergoes quantum phase transition. The formation of excitations close to the critical point due to the divergence of relaxation time results in the loss of adiabaticity, thus reducing the performance of the quantum machine \cite{dutta15quantum, RevModPhys.83.863, dziarmaga10dynamics}.  Although conventional control techniques such as STA involve complex calculations and non-trivial many-body interactions, we propose the implementation of the {\it bath-engineered quantum  engine} (BEQE), in which the working of the engine can be improved significantly through the simple control of bath spectral functions. 

The present work is organized as follows. We describe a many-body quantum Otto cycle in Section \ref{secII}. In Section \ref{secIII}, the operation of the BEQE using a generic free fermionic WM is explained in detail. We study a specific example of the BEQE using the transverse Ising model in Section \ref{secIV}. Finally, we summarize our results in Section \ref{secV}. Details of the calculations presented in this work are included in the Appendix. 

\section{Many-Body Quantum Otto Cycle} \label{secII}

\begin{figure}[h]
\includegraphics[width = 9cm]{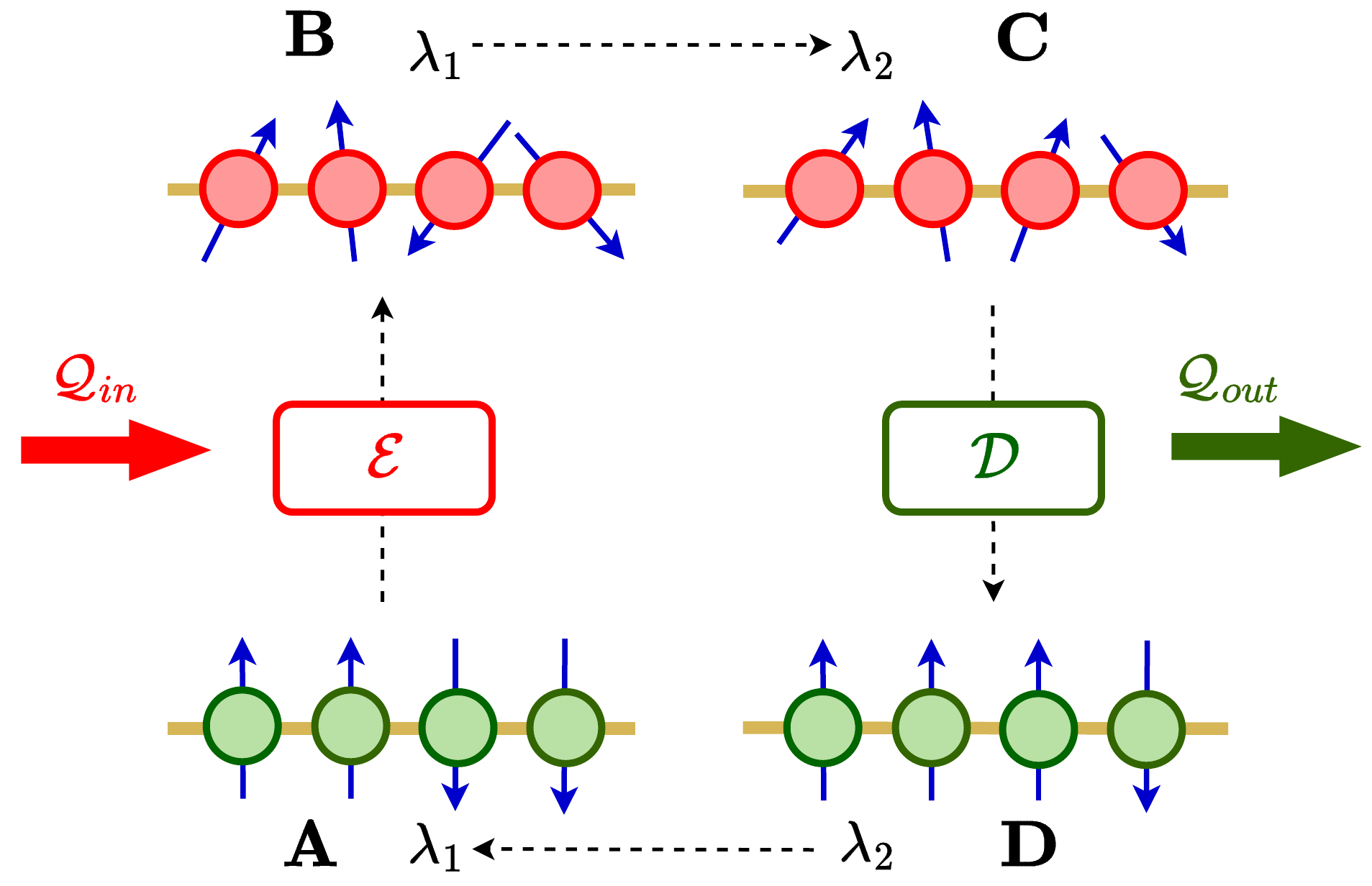}
\caption{Schematic diagram of a quantum Otto cycle with a many-body spin system as the working medium.}
\label{fig_cycle}
\end{figure}

We consider an Otto cycle with the working medium (WM) described by the Hamiltonian $H(\lambda (t))$, where $\lambda$ is a time-dependent parameter. The four-stroke quantum Otto cycle consists of two non-unitary strokes and two unitary strokes, as described below (Figure \ref{fig_cycle}).\vspace{-6pt}

\begin{itemize}
\item[(i)]{{Non-unitary stroke} {A $\rightarrow$ B}}: The 
 WM with parameter $\lambda = \lambda_{1}$ is connected to an energizing bath $\mathcal{E}$ until  it reaches the corresponding steady state at {B} by receiving energy $\mathcal{Q}_{in}$ from the bath.

\item[(ii)]{{Unitary stroke} {B $\rightarrow$ C}}: The WM is decoupled from the energizing bath and  $\lambda$ is changed from $\lambda_{1}$ to $\lambda_{2}$ at a speed of $1/ \tau_1$.

This unitary evolution is described by the Von Neumann equation of motion:
\begin{equation}
\frac{d\rho}{dt} = -i [H, \rho].
\end{equation}

\item[(iii)]{{Non-unitary stroke} {C $\rightarrow$ D}}: The WM with $\lambda = \lambda_{2}$ is now connected to a decaying bath $\mathcal{D}$  until it reaches the corresponding steady state at {D}; energy $\mathcal{Q}_{out}$ flows from the WM to the bath during this stroke.

\item[(iv)]{{Unitary stroke} {D $\rightarrow$ A}}: After decoupling from the decaying bath, the parameter $\lambda$ is changed back to $\lambda_{1}$ from $\lambda_{2}$
with a speed of $1/ \tau_2$.

\end{itemize} 

{{The WM crosses a quantum critical point at $\lambda = \lambda_c$ during the unitary strokes, such that $\lambda_2 \leq \lambda_c < \lambda_1$.}} The energy at the end of stroke $i$ is calculated using the equation
\begin{equation}
{E}_i = \text{Tr} (H^i \rho^i)
\end{equation}
where $H^i$ and $\rho^i$ are the Hamiltonian and the density matrix at $i = {{A}}, {{B}}, {{C}}, {{D}}$.
The heat input ($\mathcal{Q}_{in}$) and heat output ($\mathcal{Q}_{out}$) can be calculated using
\begin{eqnarray}
\mathcal{Q}_{in} &=& {E}_B - {E}_{A}\\
\mathcal{Q}_{out} &=& {E}_D - {E}_{C}.
\end{eqnarray}

The output work is given by $\mathcal{W} = - (\mathcal{Q}_{in} + \mathcal{Q}_{out})$. The sign convention used here is as follows: energy is taken to be positive (negative) if it enters (leaves) the WM.
The Otto cycle works as an engine when $\mathcal{Q}_{in} >0, \mathcal{Q}_{out} < 0$ and $\mathcal{|W|} < 0$.
The performance of the engine is then characterized using the quantity of efficiency ($\eta$), which is defined as
\begin{eqnarray}
\eta &=& -\frac{\mathcal{W}}{\mathcal{Q}_{in}}.
\end{eqnarray}
{We note that other regimes of operation  may arise for different signs of $\mathcal{Q}_{in}$ and $\mathcal{Q}_{out}$, as discussed in \protect{\cite{PhysRevE.94.062109}}.}

\section{Bath Spectral Form Engineering}
\label{secIII}

We consider a free fermionic WM, described by a Hamiltonian of the form
\ba
H = \sum_k \psi_k^{\dagger} H_k \psi_k \non \\
H_k = \vec{f}(k).\vec{\sigma}_k
\ea
where $\vec{\sigma}_k = \left(\sigma_k^x, \sigma_k^y, \sigma_k^z \right)$ denotes the Pauli matrices corresponding to the $k$-th mode; $\vec{f}(k)$ is a model-dependent function for the $k$-th mode; and $\psi_k^{\dagger} = \left(c_{1k}^{\dagger}~c_{2k}^{\dagger}\right)$, where $c_{jk}$ and $c_{jk}^{\dagger}$ (with $j = 1, 2$) denote the fermionic operators corresponding to the $k$-th mode.

For non-interacting $k$ modes, the density matrix $\rho$ of the system can be written as $\rho = \otimes_k \rho_k$. The WM undergoes unitary dynamics during the strokes D $\to$ A and {B} $\to$ {C}, described by the Von Neumann equation: 
\ba
\dot{\rho}_k = -i \left[H_k, \rho_k \right]
\label{eqVN}
\ea
for each $k$ mode.
Furthermore, we assume fermionic baths such that each $k$ mode evolves independently during the non-unitary strokes, described by the master equation \cite{keck17dissipation}
\ba
\frac{d \rho_k}{dt} =  \mathcal{G}_{\alpha}(\Delta_k) \mathcal{L}_k \left[\rho_k(t)\right] + \mathcal{G}_{\alpha}(-\Delta_k)\mathcal{L}_k^{\dagger} \left[\rho_k(t)\right]
\label{eqDiss}
\ea
where, following the Kubo--Martin--Schwinger condition, {{we have
\ba
\mathcal{G}_{\alpha}(-\Delta_k) = \exp(-\Delta_k/ T_{\alpha}) \mathcal{G}_{\alpha}(\Delta_k).
\label{eq:kms}
\ea}}

Here $\mathcal{G}_{\alpha}(\nu)$ denotes the spectral function of the  $\alpha = \mathcal{E}, \mathcal{D}$ bath at frequency $\nu$, whereas $T_{\alpha}$ is the effective temperature of the $\alpha$ bath \cite{breuer02, revathy20universal}. The superoperator $\mathcal{L}_k$ {{and $\mathcal{L}_{k}^{\dagger}$ are}} of the form
{{
\ba 
\mathcal{L}_k &=& \left( L_{k} \rho_{k}
L_{k}^{\dagger} -\frac{1}{2}\lbrace L_{k}^{\dagger}L_{k}, \rho_k \rbrace \right) \label{eq_stroke1}\\
\mathcal{L}_k^\dagger &=& \left( L_{k}^\dagger \rho_{k}
L_{k} -\frac{1}{2}\lbrace L_{k}L_{k}^{\dagger}, \rho_k \rbrace \right)
\non
\ea}}
with $L_k$ being the Lindblad operators denoting jumps between the different eigenenergy levels.
The above dynamics given in {{Equations \eqref{eqDiss}--\eqref{eq_stroke1}}} ensures that each $k$ mode thermalizes independently with the bath, such that the steady state of the WM at the end of an isochoric stroke is given by $\rho = \otimes_k \rho_k^{\rm th}$, where $\rho_k^{\rm th}$ is the Gibbs state corresponding to the $k$-th mode \cite{PhysRevB.83.094304}. {{However, in general, $\Delta_k$ can be expected to vary with $k$, which may result in the global state $\rho = \otimes_k \rho_k^{\rm th}$  being non-thermal.
We emphasize that even though globally the WM remains in a non-thermal state, in contrast to quantum engines powered by squeezed thermal baths,  none of the $k$ modes receive any ergotropy from the $\mathcal{E}$ and $\mathcal{D}$ baths in this setup \cite{PhysRevLett.112.030602, niedenzu2018quantum}. Furthermore, this global steady state becomes thermal for $\Delta_k$ becoming a positive $k$-independent  constant, which may happen far away from a quantum critical point, or in the limits $T_{\alpha} \to \infty$ and $T_{\alpha} \to 0$. In addition, we note that the bath considered here is local in $k$ space, and therefore can be expected to be non-local in real space. However, this bath can become local far away from the critical point, in which regime the system may be composed of effectively non-interacting particles (see Section {\ref{secIV}}). A detailed discussion regarding the bath considered here is given in \cite{keck17dissipation}.}}

Non-adiabatic excitations are inevitable when a quantum system is driven across quantum critical points \cite{dutta15quantum}. This results in a reduction in output work as well as efficiency when a quantum critical substance is made the working medium of a quantum Otto engine~\cite{revathy20universal}. Here, we propose  bath spectral form engineering to prevent these excitations from reducing the performance of the engine, henceforth called the \textit{bath-engineered quantum engine} (BEQE).

\begin{figure}[h]

\includegraphics[width=8cm]{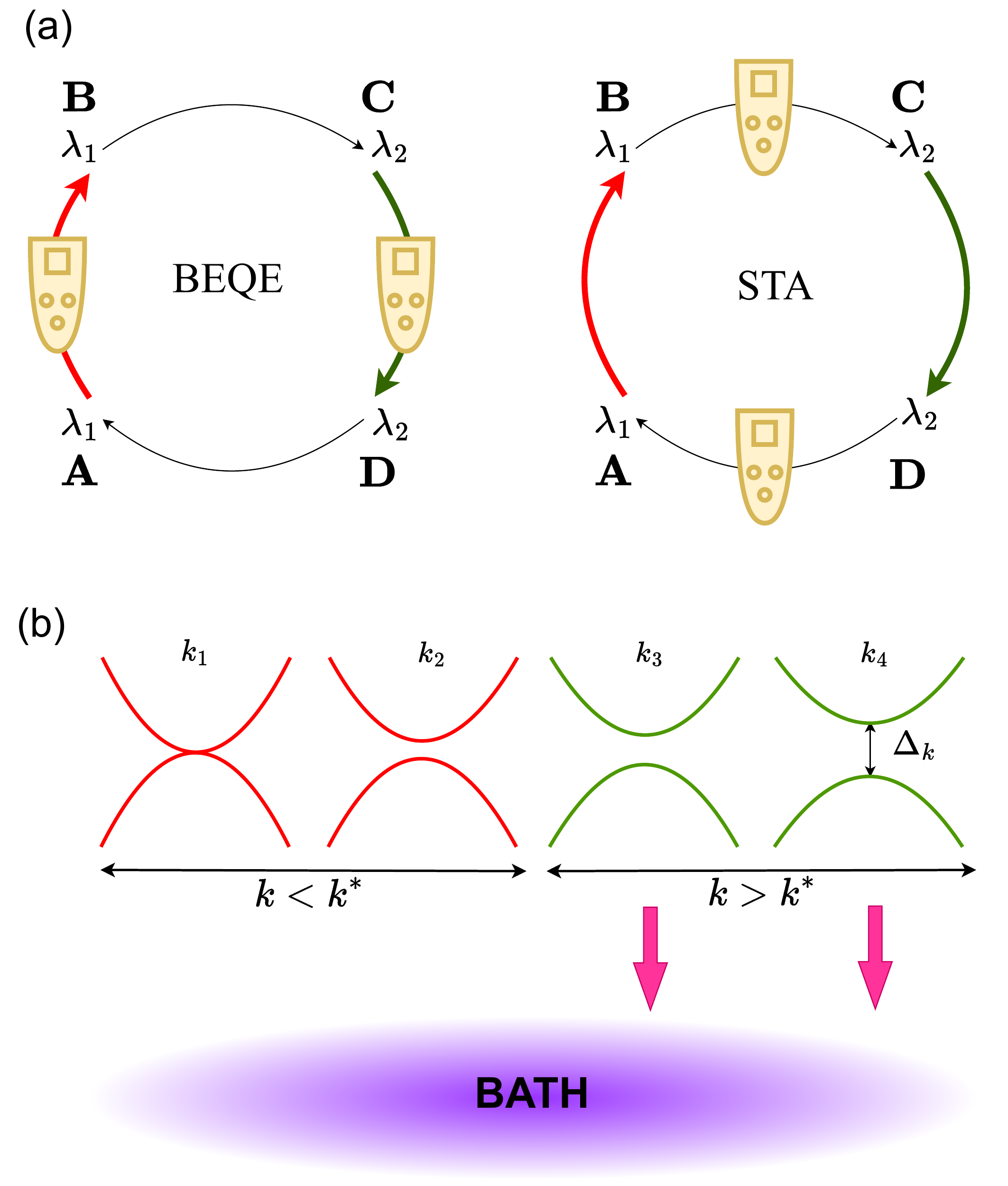}
\caption{(\textbf{a}) Schematic diagram showing bath engineering being applied during the non-unitary strokes, whereas shortcuts to adiabaticity are applied during the unitary strokes. (\textbf{b}) A $k$ mode is coupled to the bath if $ \Delta_k > \Delta^*$ ($k>k^*$) and is not coupled to the bath if $ \Delta_k < \Delta^*$ ($k<k^*$), where $\Delta_{k^*} = \Delta^*$. }\label{fig_2}
\end{figure}

In the bath engineering technique, we choose bath spectral functions $\mathcal{G}_{\alpha}$ with appropriate cut-offs, such that the modes which have higher probabilities of getting excited, and which are therefore detrimental to the performance of the finite-time quantum engine, are not allowed to participate in the dynamics.
Although techniques such as shortcuts to adiabaticity are applied in the unitary strokes, bath engineering is performed during the non-unitary strokes (Figure \ref{fig_2}a).

The physics of excitations generated in a system which is driven at a finite rate across a quantum critical point (QCP)  are well established and are described by the Kibble--Zurek mechanism (KZM) \cite{zurek05dynamics, polkovnikov05universal, dziarmaga10dynamics}, {{which was also experimentally demonstrated in {\cite{cui20experimentally, PhysRevResearch.2.033369}}}}.  According to the adiabatic-impulse approximation \cite{damski06adiabatic}, these excitations occur due to vanishing energy gaps which are below a threshold value (say, $\Delta^*$), the expression of which can be obtained using KZ arguments as described below. Bath engineering is carried out such that 
the energy levels having gaps less than $\Delta^*$ are not allowed to interact with the bath, i.e, $\mathcal{G}(\Delta_k < \Delta^*) \approx 0$, thus preventing them from participating in the operation of the cycle (Figure \ref{fig_2}b).

\subsection*{Kibble--Zurek-Mechanism-Assisted BEQE}

According to KZM, the response of a system driven across a quantum critical point is determined by the inherent time scale (relaxation time $\xi_{\tau}$) of the system, and the rate of change of the system Hamiltonian \cite{polkovnikov05universal, zurek05dynamics, damski06adiabatic}. When the relaxation time $\xi_{\tau}$ of the system is greater than the rate at which the Hamiltonian parameter $\lambda$ is changed, the system stops evolving adiabatically, thus resulting in non-adiabatic excitations. In order to arrive at a more quantitative analysis,
let us assume that $t^*$ is the time at which the system looses adiabaticity and excitations begin to occur.
The energy gap $\Delta_{k_c}$ at the critical mode $ k_c$  scales with the  distance from the critical point $\lambda_c$ as {{\cite{sachdev_2011}}
\begin{equation}
\Delta_{k_c} \sim |\lambda - \lambda_c|^{\nu z},
\label{eq:Deltac}
\end{equation}}
where  $\nu$ and $z$ are the correlation length and dynamical critical exponents, respectively.
When the parameter $\lambda$ is varied using the quench protocol $\lambda = \lambda_2 + (\lambda_1 - \lambda_2)\frac{t}{\tau}$, one can write
\begin{equation}
\lambda - \lambda_c = \lambda_2 - \lambda_c + (\lambda_1 - \lambda_2) \frac{t}{\tau}.
\end{equation} 
{{According to the adiabatic-impulse approximation {\cite{damski06adiabatic}}}}, the time $t^*$ is determined by the condition that the relaxation time $\xi_{\tau}$ is of the order of the time scale with which $\lambda$ is changed, i.e., 
\ba
\frac{\Delta_{k_c}}{\dot{\Delta}_{k_c}}|_{t = t^*} \sim \xi_{\tau}.
\label{eq:KZM}
\ea 
Furthermore, the relaxation time diverges according to the scaling {
\begin{equation}
\xi_\tau \sim \frac{1}{\Delta_{k_c}} \sim |\lambda-\lambda_c|^{-\nu z}.
\label{eq:xi}
\end{equation}}
Using  the expressions \eqref{eq:Deltac}--\eqref{eq:xi} one obtains
{{
\begin{eqnarray}
\frac{\Delta_{k_c}}{\dot{\Delta}_{k_c}}|_{t = t^*} &\sim& \frac{(\lambda_2 - \lambda_c) + (\lambda_1 - \lambda_2) \frac{t^*}{\tau}}{\nu z (\frac{\lambda_1 - \lambda_2}{\tau})}\non\\
&\sim& [(\lambda_2 - \lambda_c) + (\lambda_1 + \lambda_2) \frac{t^*}{\tau}]^{-\nu z}\\
\Rightarrow t^* &\sim & t_c + \frac{\tau}{\lambda_1 - \lambda_2}\left( \nu z (\frac{\lambda_1 - \lambda_2}{\tau}) \right)^{\frac{1}{1+ \nu z}}
\end{eqnarray}
where $t_c =  \tau(\lambda_c - \lambda_2)/\left(\lambda_1 - \lambda_2\right)$
is the time such that $\lambda(t_c) = \lambda_c$, and we have assumed that $\left(\lambda - \lambda_c\right) > 0$ for simplicity.}}

Thus, the energy gap at which the excitations begin to happen for the critical mode is given by
{{
\begin{equation}
\tilde{\Delta}^* = \Delta_{k_c}|_{t^*} \sim \left(\lambda(t^*) - \lambda_c \right)^{\nu z} \sim \left( \frac{\nu z (\lambda_1 - \lambda_2)}{\tau}\right)^{\frac{\nu z}{1 + \nu z}}.
\label{deltacr}
\end{equation}}}
{{In the quantum Otto cycle, bath engineering is implemented during the non-unitary strokes ${C} \to {D}$ and ${A} \to {B}$ by choosing 
\begin{eqnarray}\label{BEeq}
\mathcal{G}_{\mathcal D}(\Delta_k) &\approx 0&~~ \text{for}~~\Delta_k < \Delta^{*} \non\\
\mathcal{G}_{\mathcal E}(\Delta_k) &\approx 0&~~ \text{for}~~\Delta_k < \gamma \Delta^{*}\non\\
\Delta^{*} &=& C_1 \tilde{\Delta}^*
\end{eqnarray}
respectively, such that small energy gaps which have a higher probability of getting excited do not participate in the dynamics. The scaling parameter $C_1$ ($C_1 > 0$), along with \mbox{Equations \eqref{deltacr} and \eqref{BEeq}}, determine the lower cut-offs for the bath spectral functions; one can choose an appropriate $C_1$ depending on the details of the setup and the constraints involved in order to improve the performance of an engine. 
In the numerical results given below, we have chosen $C_1 = 1$ for simplicity;}} $\gamma$ is the scaling factor by which a typical energy gap changes in the ${D} \to {A}$ stroke, {{and we have assumed that $\lambda_2$ is close to the quantum critical point, i.e., 
\ba
|\lambda_2 - \lambda_c|^{\nu z} \ll  \tilde{\Delta}^*.
\ea
For non-critical $\lambda_2$ (i.e., $|\lambda_2 - \lambda_c|^{\nu z} \gg \tilde{\Delta}^*$), the energy gaps of the system at $\lambda_2$ are of the order of 
\ba
\Delta_k \approx C_2\left|\lambda_2 - \lambda_c \right|^{\nu z} + f\left(k, h_2\right),
\label{eqenergynonc}
\ea
where $C_2$ is a model-dependent constant related to the minimum energy gap of the system, whereas $f(k, h_2)$ is a model-dependent function for the mode $k$. For low-energy modes, one can expect $|f(k)| \ll C_2\left|\lambda_2 - \lambda_c \right|^{\nu z}$ \cite{sachdev_2011}. Consequently, in this case we take 
\ba
\mathcal{G}_{\mathcal D}(\Delta_k) &\approx 0~~& \text{for}~~\Delta_k < \Delta^{*} \non\\
\mathcal{G}_{\mathcal E}(\Delta_k) &\approx 0~~& \text{for}~~\Delta_k < \gamma \Delta^{*} \non\\
\Delta^* &=& C_2 \left|\lambda_2 - \lambda_c \right|^{\nu z} + C_3.
\label{eqbeqnonc}
\ea
As before, $C_3$  ($|C_3| \ll C_2 \left|\lambda_2 - \lambda_c \right|^{\nu z}$) is a constant which we choose depending on the details of the WM and constraints on the bath spectral functions. We note that ideally one should consider $C_3$ to be a function of $\tau$; however, in contrast to Equations \eqref{deltacr}  and \eqref{BEeq}, here we consider a $\tau$-independent 
  $\Delta^*$ since $C_3$ can be considered to be a small correction over the first term $C_2 \left|\lambda_2 - \lambda_c \right|^{\nu z}$ (see Equation \eqref{eqenergynonc} and the text below).

In this control protocol, the bath spectral functions of the modes with large $\Delta_k$ (see Equations \eqref{BEeq} and \eqref{eqbeqnonc}), and therefore the thermalization times for these modes, remain unchanged and finite. On the other hand, the modes with small $\Delta_k$ do not evolve during the non-unitary strokes. Consequently, the durations of the non-unitary strokes of a BEQE, and in turn the total cycle period, remain the same as that of a finite-time engine without controls. Furthermore,  only the modes with large $\Delta_k$ values go to their respective steady states at the end of a non-unitary stroke  in a BEQE, thereby in general giving rise to non-thermal global steady states at $B$ and $D$.}}

Next we demonstrate the bath engineering technique using a free fermionic model, which is described in the following section.

\section{BEQE with Transverse Ising Model as a WM}\label{secIV}

A prototypical example of a free fermionic system undergoing quantum phase transition is the one-dimensional transverse Ising model (TIM). It is an exactly solvable model and is thus widely studied. 
The Hamiltonian of the transverse Ising model is
\begin{equation}\label{H}
H(t) = -J\sum_{n} \sigma_{n}^{x}\sigma_{n+1}^{x} - h(t) \sum_{n} \sigma_{n}^{z}
\end{equation} 
where $J$ is the nearest neighbor interaction strength; $h(t)$ is the transverse field which is time-dependent, playing the role of $\lambda$ in the previous section; and $n$ is the lattice site index. Here, $\sigma_{n}^i$ with $i = x, y, z$ are the Pauli matrices at each site $n$. This system shows a zero temperature 
 quantum phase transition from a paramagnetic to ferromagnetic state at the quantum critical point $h = \pm J$ \cite{lieb61two, pfeuty70the, bunder99effect}. {{We set $J = 1$ throughout the paper so that $h = \pm 1$ are the critical points.}}

After performing Jordan--Wigner fermionization and taking the Fourier transform, the Hamiltonian $H_k$ takes the form \cite{pfeuty70the}:
\ba
H_k = -2\left(h - \cos k \right) \sigma^z + 2\sin k \sigma^x.
\ea
Even though unitary dynamics allows transitions only between $\ket{0}$ and $\ket{k, -k} = c_k^{\dagger} c_{-k}^{\dagger} \ket{0}$, the system bath interactions lead to transitions to the $\ket{\pm k} = c_{\pm k}^{\dagger}\ket{0}$ states as well, resulting in the mixing of states \cite{keck17dissipation, bandyopadhyay18exploring}. Therefore, the Hamiltonian is rewritten in the basis $|0\rangle, |k\rangle, |-k\rangle, |k, -k\rangle$ as

\begin{equation}
H_{k} = \begin{bmatrix}
 -2(h - \cos k) & 0 & 0 & 2 \sin k \\ 0 & 0 & 0 & 0\\ 0 & 0 & 0 & 0\\ 2 \sin k & 0 & 0 & 2(h - \cos k) 
\end{bmatrix}
\label{eqHk}
\end{equation}
with eigenenergies $-\epsilon_k, 0, 0, \epsilon_k$ where $\epsilon_k = 2\sqrt{(h - \cos k)^{2} + {\sin k}^{2}}$.

We now focus on the strokes of the Otto cycle with the TIM as the WM. The density matrix at {B} is given by $\rho^B$  ($=\otimes_k \rho_k^{B}$), where 
\begin{equation}\label{eqn_rhoB}
\rho_{k}^B = \begin{bmatrix}
 \frac{e^{\beta \epsilon_k}}{Z_k} & 0 & 0 & 0 \\ 0 & \frac{1}{Z_k} & 0 & 0\\ 0 & 0 & \frac{1}{Z_k} & 0\\ 0 & 0 & 0 & \frac{e^{- \beta \epsilon_k}}{Z_k}
\end{bmatrix}
\end{equation} 
is the thermal state for the mode $k$ corresponding to $T = T_H$ and $h = h_1$.

Here $\beta = \frac{1}{k_{B} T_H}$ ($k_B$ is set to unity for the rest of the paper) and $Z_k = 2 + e^{\beta \epsilon_k} + e^{-\beta \epsilon_k}$ are the partition functions for each $k$ mode.
In the unitary stroke ({B} $\rightarrow$ {C}), the transverse field is changed from $h_1$ to $h_2$ according to the protocol,
\begin{equation}
h(t) = h_{1} + (h_{2} - h_{1})(\frac{t}{\tau_1}), \hspace{1cm} t \in [0, \tau_1]
\end{equation}
in a time $\tau_1$ with $h_1 \gg h_2$.
During the non-unitary stroke ({C} $\rightarrow$ {D}) the system again reaches a state  $\rho^D =\otimes_k \rho_k^{D}$, where $\rho_k^{D}$ is the thermal state for the mode $k$ corresponding to $T_C$ and $h = h_2$ at {D}. The transverse field $h_2$ is then changed back to $h_1$ using the same quench protocol but in time $\tau_2$ in the unitary stroke {D} $\rightarrow$ {A}.

Now let us examine how bath engineering is implemented in TIM.  {{We first focus on the case of $h_2 \to 1$}.}
As discussed before, we make use of selective coupling between the bath and the working medium so that some $k$ modes close to the critical mode $k_c$, having an energy gap $\Delta_k$ lower than  the threshold value $\Delta{^*}$ ($= \left( \frac{\nu z (h_1 - h_2)}{\tau}\right)^{\frac{\nu z}{1 + \nu z}}$), {{are}} prohibited from interacting with the bath, thereby preventing these modes from thermalizing.\\

{{The energy gaps between the adjacent non-degenerate eigenstates of the Hamiltonian \eqref{eqHk} are given by
\ba
\Delta_k =  \epsilon_k 
= 2\sqrt{(h-\cos k)^2 + \sin^2 k}.
\label{eqdelk}
\ea}}
For TIM, the critical exponents are $\nu = 1$ and $ z = 1$ so that $\Delta^*$, as obtained in Equations (\ref{deltacr}) and \eqref{BEeq}, is given by 
\begin{equation}
\Delta^* = \sqrt{\frac{h_1 - h_2}{\tau_2}}
\label{eq:DeltastarTIM}
\end{equation}
Below we present the steps required to incorporate bath engineering into the quantum Otto cycle.
\begin{itemize}
\item[(a)] As discussed above, we consider a lower cut-off for the decaying bath spectral function, given by $\mathcal{G}_{\mathcal{D}}\left(\Delta\right) \approx 0$ for   $\Delta_k < \Delta^*$ (see Equations \eqref{BEeq} and \eqref{eq:DeltastarTIM}).
This choice of bath spectral function  ensures that modes with $\Delta_k < \Delta^*$ are not allowed to interact with the decaying bath, so that $\rho_k^D = \rho_k^C$ for these modes.

On the other hand, modes with $\Delta_k > \Delta^*$ thermalize with the decaying bath and reach the state $\rho_k^D = \frac{e^{-\beta_C H_k(h_2)}}{Z_k}$ at ${D}$.
\item[(b)] In the {D} $\rightarrow $ {A} stroke,  the Hamiltonian is changed from $h_2$ to $h_1$, starting from the state $\rho_k^D$ to reach $\rho_k^A$.
\item[(c)] At {A}, the lower cutoff for the energizing bath is chosen to be $\mathcal{G}_{\mathcal{E}} \approx 0$ for $\Delta_k < \gamma \Delta^*$ where $\gamma$ is chosen in such a way that $\gamma \Delta^*$ is of the order of the lower-energy gaps for $h = h_1$ which allows for some modes to be bath-engineered in the energizing bath stroke. This results in $\rho_k^B = \rho_k^A$ for such modes. 

 The modes with $\Delta_k > \gamma \Delta^*$ are allowed to 
 interact with the energizing bath, leading the system to the steady state given by Equaion (\ref{eqn_rhoB}). 
\item[(d)] From {B} to {C}, the system is quenched and all modes evolve to reach $\rho^C$.

\end{itemize}

The total heat input and output of the system are calculated using
\begin{eqnarray}
\mathcal{Q}_{in} &=& \sum_k \mathcal{Q}_{in}^k \\
\mathcal{Q}_{out} &=& \sum_k \mathcal{Q}_{out}^k
\end{eqnarray}
{{and the work output and efficiency of the engine are obtained using 
\ba
\mathcal{W} &=& \sum_k \mathcal{W}_k\non\\
\mathcal{W}_k &=& -\left(\mathcal{Q}_{in}^k + \mathcal{Q}_{out}^k\right)\non\\
\eta &=& -\frac{\mathcal{W}}{\sum_k \mathcal{Q}_{in}^k}.
\ea
} 
As discussed in Section \ref{secII}, a mode  $k$ acts as an engine with non-zero work output for $\mathcal{Q}_{in}^k > 0$, $\mathcal{Q}_{out}^k < 0$ and $\mathcal{W}_k < 0$.}

\begin{figure}[h]

\includegraphics[width=8cm]{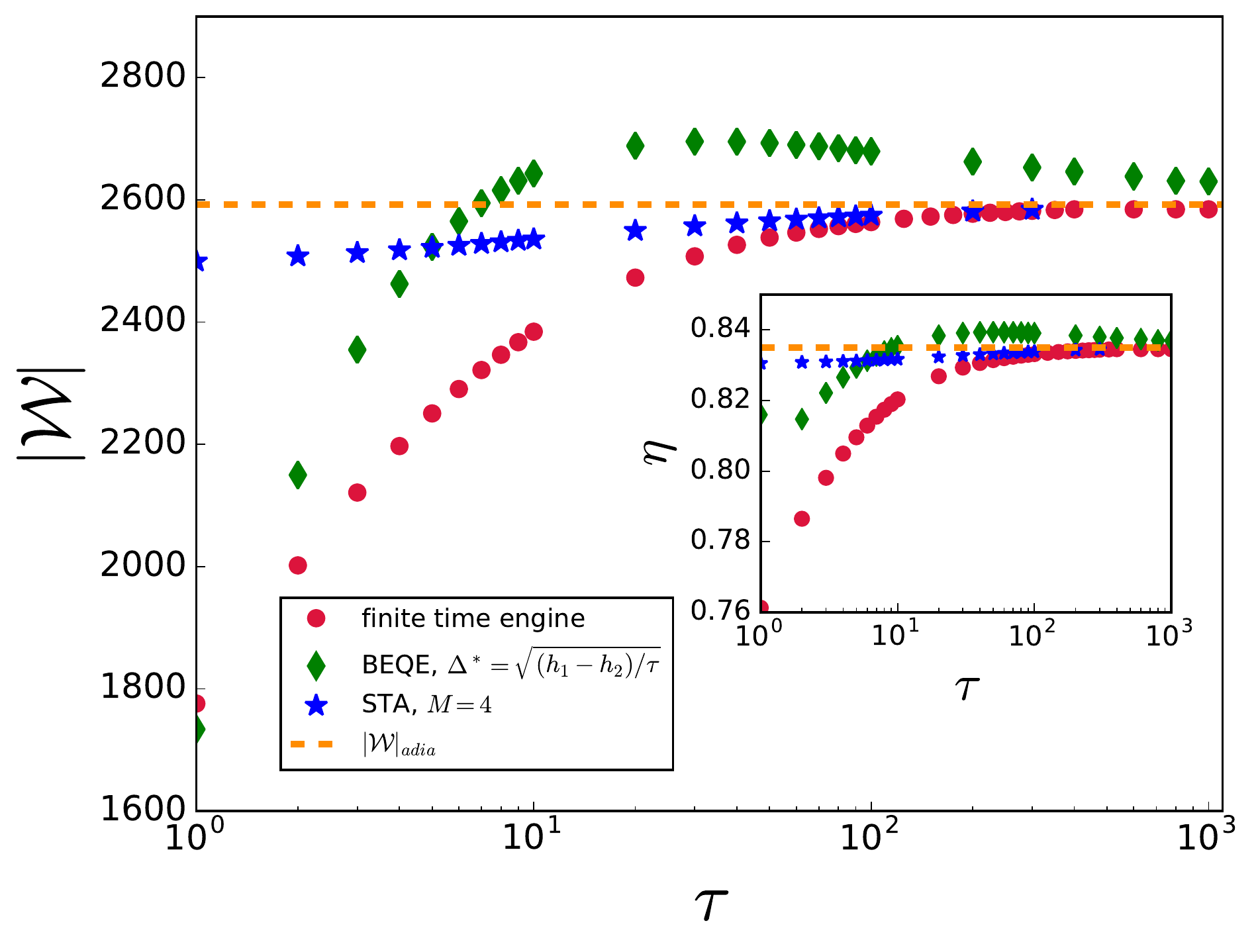}
\caption{$|\mathcal{W}|$ is plotted as a function of $\tau$ for the critical engine using different techniques. Inset: $\eta$ is plotted as a function of $\tau$. The parameters used are $L=1000, h_1 = 10, h_2 = 1, T_H = 20, T_C = 1, \gamma = 6.5, \tau_1 = \tau_2 = \tau$. 
}
\label{fig_tim_W}
\end{figure}

We {depict} the variation of the output work and the efficiency  of the engine after implementing bath engineering in Figure \ref{fig_tim_W}. To obtain a complete picture,  we compare BEQE with finite-time engines without any control, finite-time engines with the presence of shortcuts to adiabaticity in the unitary strokes, and engines operating in the adiabatic limit, i.e., $ \tau_1 = \tau_2 = \tau \rightarrow \infty$ (or infinite-time engines). As shown in \cite{campo12assisted}, the STA Hamiltonian involves long-range interactions. However, one can truncate the control Hamiltonian to $M$-body terms to obtain a physically realizable approximate STA protocol.
In Figure \ref{fig_tim_W}, we present a comparison of the output work as a function of $\tau$ ($= \tau_1 = \tau_2)$. 
As expected, engines using STA 
  always {perform} better than the finite-time engines without controls. However, interestingly, the BEQE {outperforms} the engines using STA, as well as the perfectly adiabatic engine, for a wide range of $\tau$ values, thus exhibiting the remarkable benefits offered by the bath engineering technique.

Similarly, we also  {plot} the efficiency $\eta$ as a function of $\tau$ (see inset of Figure \ref{fig_tim_W}) and  {compare} BEQE with engines operating with different techniques. Here  {also we find} that BEQE  {outperforms} all other engines for the same range of $\tau$ values as in the work output analysis.
The expressions for $\mathcal{|W|}_{adia}$ and $\eta_{adia}$ are given in Appendix \ref{appA}.

\begin{figure}[h]

\includegraphics[width=8cm]{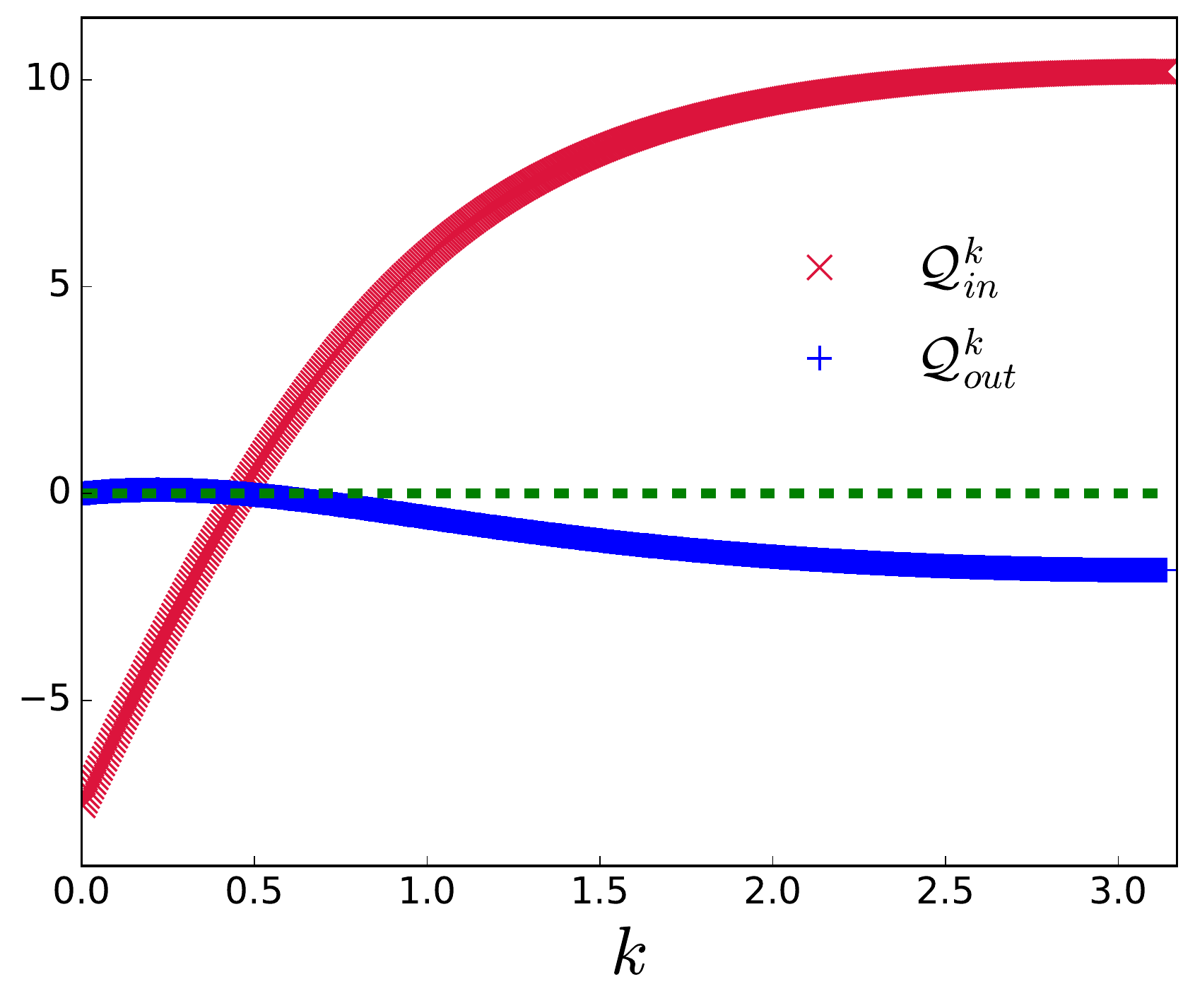}
\caption{$\mathcal{Q}_{in}^k$ and $\mathcal{Q}_{out}^k$ are plotted as functions of  $k$ modes for the critical engine in the adiabatic limit. The parameters are $L = 1000, h_1 = 10, h_2 = 1, T_H = 20, T_C = 1$. The green dashed line represents the zero of heat.
}
\label{fig_crit_adia}
\end{figure}

The fact that BEQE  {outperforms} other engines can be explained using Figure \ref{fig_crit_adia}, where the $\mathcal{Q}_{in}$ and $\mathcal{Q}_{out}$ values for a perfect adiabatic engine are plotted as a function of individual $k$ modes. In Figure \ref{fig_crit_adia}, it can be seen that even when the engine {works} in 
 the adiabatic limit, there {are} some $k$ modes close to the critical mode which {do} not function as an 'engine' ($\mathcal{Q}_{in} < 0$).  To understand this better, let us consider the adiabatic limit where $\rho_k^C = \rho_k^B$ in the eigenbasis. In this limit, $\mathcal{Q}_{in}$ for each $k$ mode is given by (see Appendix \ref{appA} for details)
\begin{align}
\mathcal{Q}_{in}^{k} &=&  \frac{\Delta_k(h_1)}{2} \Big[ \frac{\left(e^{-\frac{\beta_H \Delta_k(h_1)}{2}} - e^{ \frac{\beta_H \Delta_k(h_1)}{2}} \right) }{Z(h_1)} \\
 &-&\frac{\left(e^{- \frac{\beta_C \Delta_k(h_2)}{2}} - e^{\frac{\beta_C \Delta_k(h_2)}{2}} \right)}{Z(h_2)} \Big]
\end{align}
For $\mathcal{Q}_{in}^k$ to be positive, 
\ba \label{eq_qin_k}
\frac{\sinh(\frac{\beta_H \Delta_k(h_1)}{2})}{2 + \cosh(\frac{\beta_H \Delta_k(h_1)}{2})} < \frac{\sinh(\frac{\beta_C \Delta_k(h_2)}{2})}{2 + \cosh(\frac{\beta_C \Delta_k(h_2)}{2})}.
\ea
There can be modes for which this condition is not satisfied, resulting in 'non-engine' modes in the adiabatic limit. BEQE helps to remove these non-engine modes from participating in the non-unitary strokes, thereby boosting the performance of the engine compared to the perfectly adiabatic engine.  
It can be noted that the presence of non-engine modes is essential for the BEQE to outperform the adiabatic engine. For instance, when $T_H \rightarrow \infty$ ($\beta_H \rightarrow 0$), Equation \ref{eq_qin_k} may be satisfied for all modes so that the technique of BEQE will not provide better results compared to the adiabatic engine. However, we emphasize that although Equation \eqref{eq_qin_k} and the discussion above are specific to TIM WM, BEQE can be expected to perform better than generic finite-time free fermionic quantum critical engines, following the arguments presented in Section \ref{secIII}.

\begin{figure}[h]

\includegraphics[width=9cm]{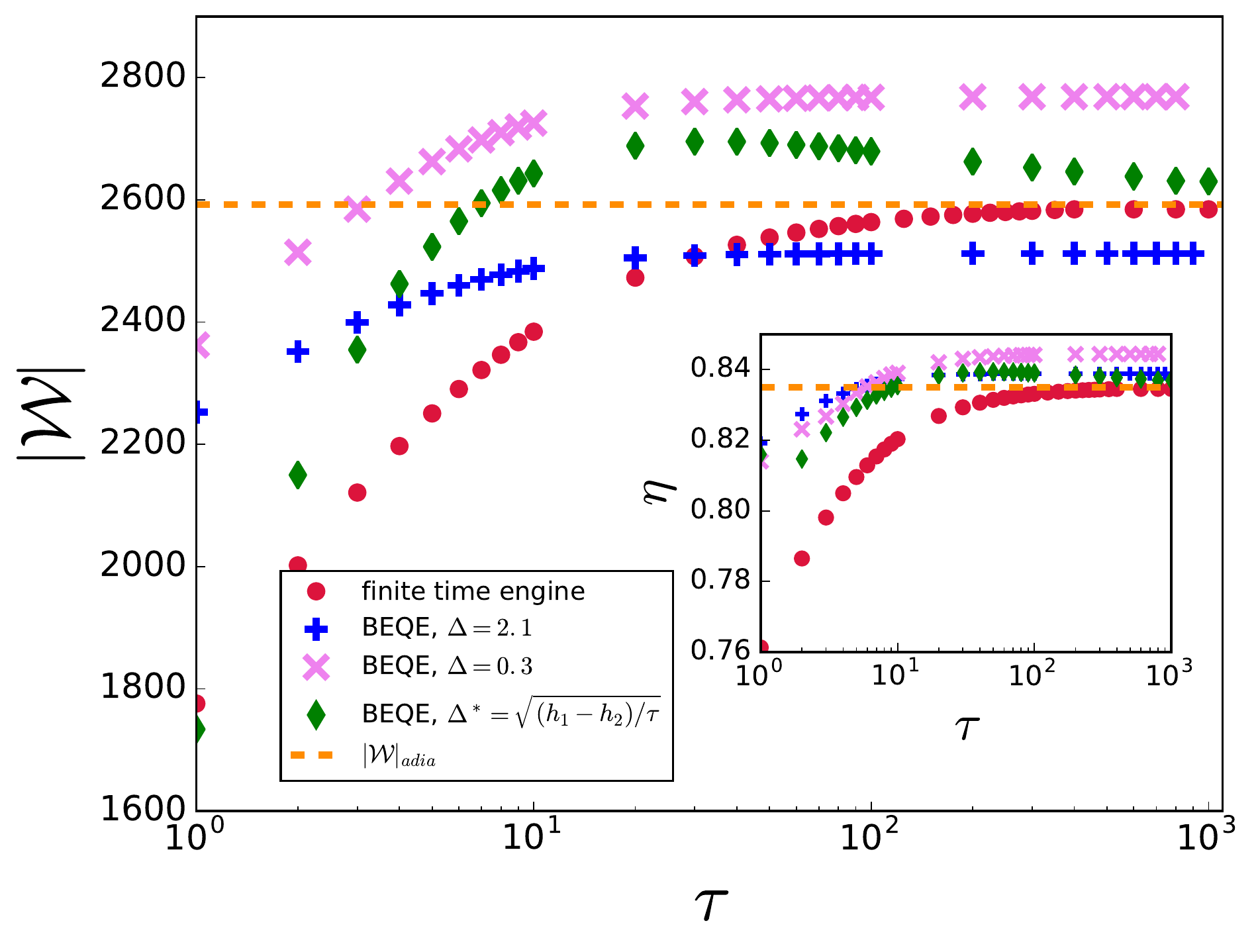}
\caption{$\mathcal{|W|}$ is plotted as a function of $\tau$ using constant value of $\Delta^*$ for all $\tau$. Inset: $\eta$ is plotted as a function of $\tau$. Here, $\gamma (\Delta^* = 2.1) = 9$ and $\gamma (\Delta^* = 0.3) = 62$. Other parameters are same as in \mbox{Figure  \ref{fig_tim_W}.} }
\label{fig_del_eta}
\end{figure}

We note that BEQE depends on the appropriate choice of bath spectral function \eqref{BEeq}, which again depends on $\tau$ through Equation \eqref{deltacr}. However, in experimental setups, it might be difficult to change the bath-spectral function for every change of $\tau$. Consequently, 
we examine the robustness of the bath engineering protocol by plotting the work output and efficiency vs $\tau$  for constant  values of $\Delta^*$. In this case also, the results show that the engine performance can be enhanced by choosing  appropriate {\it constant} values of $\Delta^*$ as shown in Figure \ref{fig_del_eta}, thus highlighting the effectiveness of the proposed protocol in practical scenarios. \vspace{-6pt}

\begin{figure}[h]

\includegraphics[width=9cm]{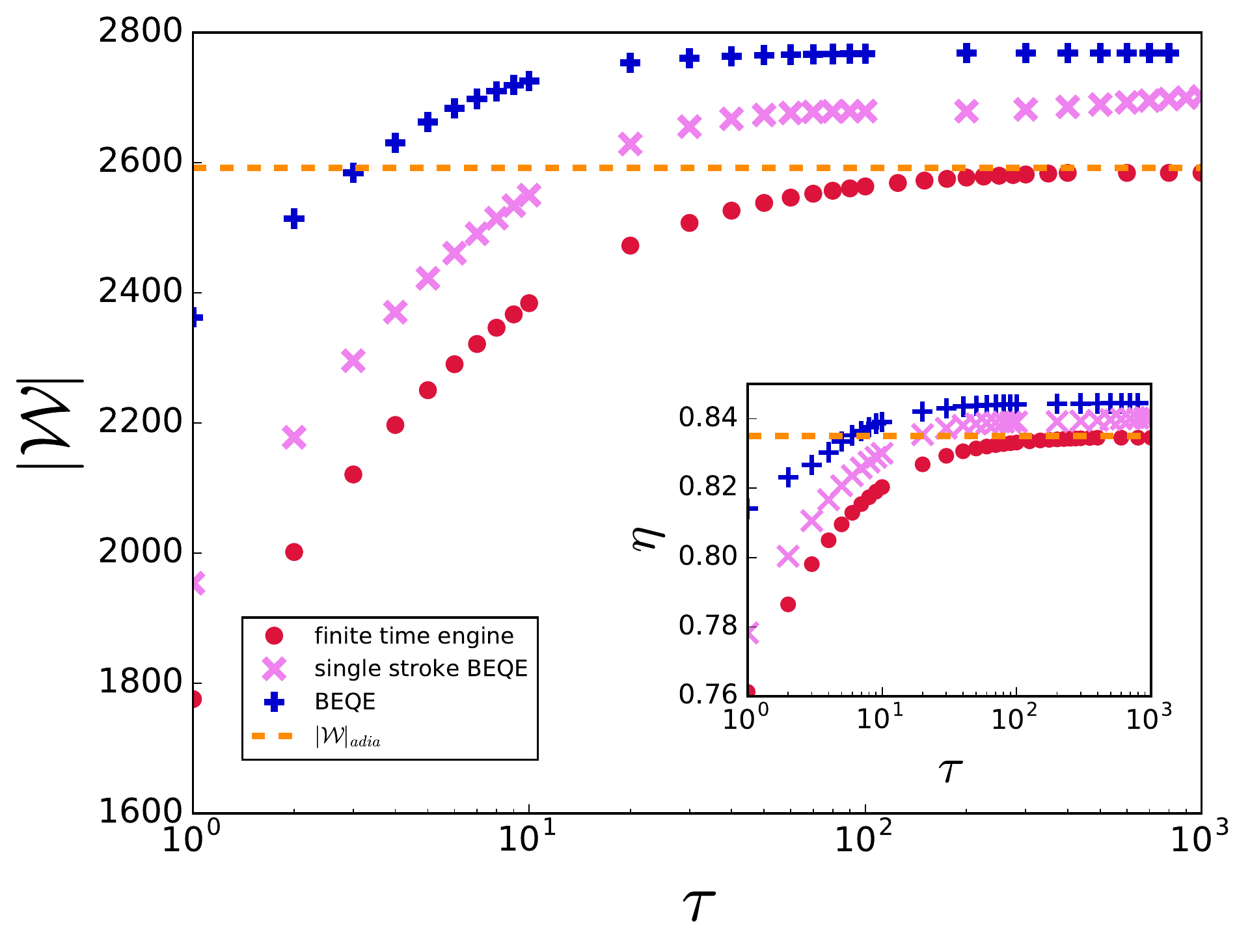}
\caption{$\mathcal{|W|}$ is plotted as a function of $\tau$ using a constant value of $\Delta^*$ for all $\tau$  with bath engineering only in the $\bf{C} \to \bf{D}$ non-unitary stroke (single-stroke BEQE) and with bath engineering in both the non-unitary strokes (BEQE). Inset: $\eta$ is plotted as a function of $\tau$. Here $\Delta^* = 0.3, \gamma = 62$. Other parameters are same as in Figure  \ref{fig_tim_W}.}
\label{fig_del_cold}
\end{figure}

We point out that one may be able to further simplify the control protocol by implementing bath engineering in only one of the non-unitary strokes (single-stroke BEQE). In Figure \ref{fig_del_cold} it can be seen that even a single-stroke BEQE performed better than the finite- and infinite-time engines. Therefore, this simplified protocol can be helpful as long as the overall work output can be increased, which one can calculate following the mechanism discussed above, even though there might be scenarios in which this simplified protocol may not suffice.

\begin{figure}[h]

\includegraphics[width=9 cm, height = 6 cm]{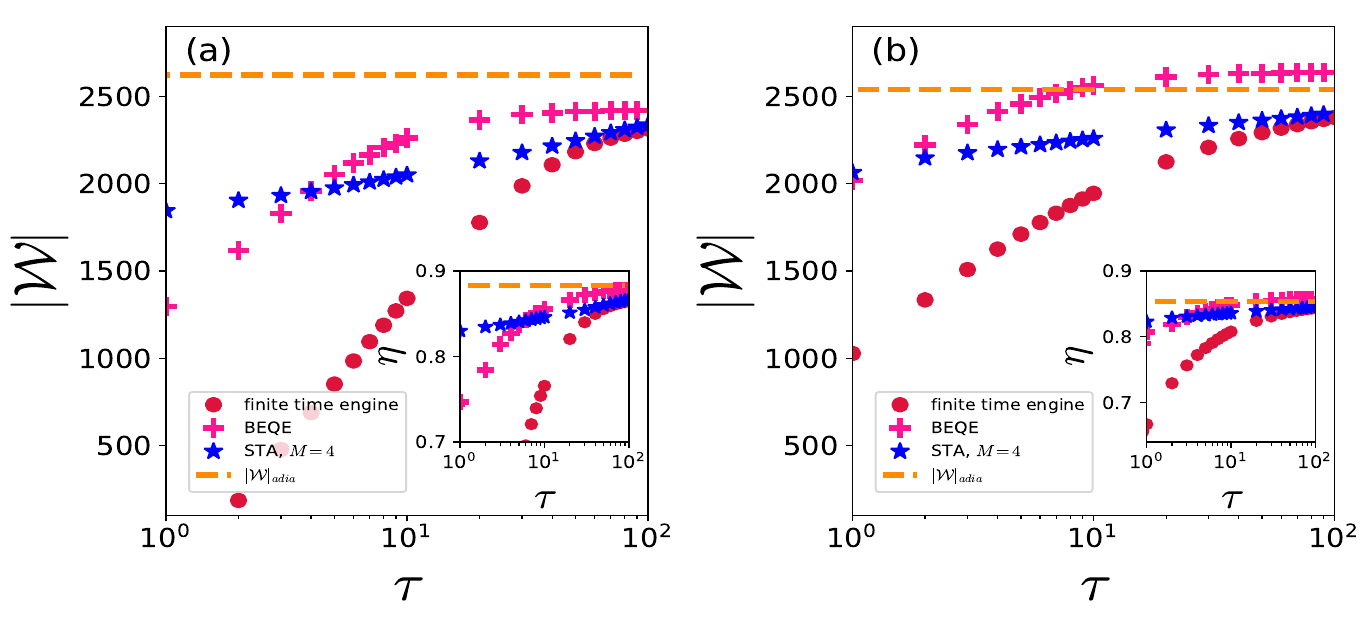}
\caption{$\mathcal{|W|}$ is plotted as a function of $\tau$ for (a) $h_2 = 0.5, C_3 = 0.02$ and (b) $h_2 = 0.8, C_3 = 0.08$, with $\Delta^*$ given by Equation \eqref{eqbeqnonc}.  Here $L = 1000, h_1 = 10,  T_H = 20, T_C = 1, \tau_1 = \tau_2 = \tau, C_2 = 2$ (see Equation \eqref{eqdelk}). }
\label{fig_otherh2}
\end{figure}

{{Even though we have set $h_2 = 1$ for Figures \ref{fig_tim_W}, \ref{fig_del_eta}, and \ref{fig_del_cold}, the improvement shown by BEQE persists when one crosses the quantum critical point during the unitary strokes. This is shown in Figure \ref{fig_otherh2} in which we have used  (see Equation \eqref{eqbeqnonc})
\ba
\Delta^* = C_2 (1 - h_2) + C_3
\ea
to improve the output work and efficiency for $h_2 < 1$.

We note that as we decrease $\tau$, more $k$ modes get excited and become detrimental to the performance of the engine, thereby
resulting in a diminishing work output {in all cases,} in the small $\tau$ regime. This, in
turn, is also reflected in the decreasing power output for small
values of $\tau$ , as shown in Figures \ref{fig_power} and \ref{fig_powernc} in Appendix  \ref{appB}.}}

%

\section{Conclusions}\label{secV}

We have proposed the bath-engineered quantum engine, in which, through the appropriate choice of bath spectral functions, one can dramatically boost the performance of quantum critical engines based on free fermionic WMs. The operation of the BEQE inherently depends on the Kibble--Zurek mechanism; consequently, knowledge about the universality class of the WM and the strokes of the Otto cycle suffice to implement this method. This is in stark contrast to conventional methods such as shortcuts to adiabaticity, in which one may need detailed knowledge about the eigenspectrum for their application. We emphasize that the improvement in performance in the case of the BEQE is also accompanied by  {{the simplicity}}    of formulating the control protocol, as compared to more conventional techniques, such as STA, {{in which the unitary strokes may}} involve non-trivial many-body terms in quantum systems driven through quantum critical points \cite{campo12assisted, kolodrubetz17geometry, sels17minimizing}.
{{We note that the costs of applying STA and the BEQE also raise important questions. The cost of STA has been widely studied in the literature, for example, in \cite{PhysRevLett.118.100601, PhysRevE.99.022110, PhysRevA.96.022133, RevModPhys.91.045001}. These costs depend on the control Hamiltonian, which in the case of critical systems may involve long-range interactions {\cite{campo12assisted}} and can be expected to be strongly dependent on the details of the setup in question. In the case of the BEQE, the control approach involves introducing lower cut-offs in the bath spectral functions (see Equations {\eqref{BEeq} and \eqref{eqbeqnonc}}). As discussed in \cite{PhysRevResearch.2.033285, PhysRevE.87.012140}, one can engineer these cut-offs by introducing filters in the form of harmonic oscillators with appropriate frequencies. As for STA, we expect the cost of the BEQE to depend on the details of the implementation. However, a detailed discussion regarding the comparison of costs of BEQE and STA goes beyond the scope of the current paper.}}
Furthermore, interestingly, in spite of the simplicity of the proposed control protocol, our analysis with TIM WM shows that the BEQE can outperform quantum engines assisted through STA, and even infinite-time quantum engines, thus highlighting the significant benefits offered by this control method. We also exhibit the robustness of the BEQE control protocol by considering constant values of $\Delta^*$.

{{The control protocol proposed here can be expected to be most relevant in quantum critical engines, owing to the high probability of excitations in systems driven through quantum phase transitions, and the presence of many-body interaction terms in the corresponding STA Hamiltonians. However, one can expect the BEQE, which depends on introducing lower cut-offs into bath spectral functions, to be applicable even away from criticality, for multi-level WMs involving non-equispaced energy levels. On the other hand, as one can infer from Equations \eqref{BEeq} and  \eqref{eqbeqnonc}, this protocol becomes invalid in the presence of equispaced energy levels.}}

Several existing setups can be suitable for the experimental realization of the BEQE, such as trapped ions \cite{rossnagel16a,  ulm2013observation, Pyka13, Maslennikov17, PhysRevLett.123.080602}, optical lattices \cite{schreiber15observation}, superconducting qubits, nitrogen vacancy centers in diamond \cite{klatzow19experimental}, NMR qubit systems \cite{Peterson18}, etc.  For example, quantum simulators based on trapped ions have already been used to study the Kibble--Zurek mechanism in momentum space \cite{cui20experimentally}. 

Finally, we note that although this technique appears to be highly successful in the case of free fermionic WMs, open questions remain in the case of its application with non-integrable WMs, where such non-interacting $k$ modes may not exist. For example, one can choose the WM to be the antiferromagnetic transverse Ising model with a longitudinal field (LTIM), described by the Hamiltonian
\begin{equation}
H = J \sum_i \sigma_{i}^{z} \sigma_{i+1}^z - B_x(t) \sum_i \sigma_{i}^x - B_z \sum_i \sigma_{i}^z.
\end{equation}
Here $J$ is the strength of antiferromagnetic interaction, $B_z$ is a longitudinal field, and $B_x$ denotes a time-dependent transverse field. The competition between $J$ and  $B_z$ leads to a quantum phase transition from the antiferromagnetic state to the paramagnetic state at a critical value of $B_x^c$ for a fixed value of $B_z$ \cite{PhysRevB.92.104306, de2019ground}. One can model an Otto cycle using LTIM WM and implement the BEQE as described above (see {Appendix \ref{appC}})
. However, preliminary studies suggest that, unlike the case of the integrable model, there is no improvement in the output of the BEQE in this case (see Figure \ref{fig_ltim_eta}). This can be attributed to the absence of non-interacting momentum modes, as obtained for free fermionic systems. However, additional rigorous studies are needed to acquire a deeper understanding of the possibility of the application of BEQE to quantum engines based on more generic non-integrable~WMs.

\acknowledgments{R.B.S. and U.D. acknowledge CHANDRA super cluster of IIT Palakkad, in which all numerical simulations were performed. V.M. acknowledges support
from Science and Engineering Research Board (SERB)
through MATRICS (Project No. MTR/2021/000055)
and Seed Grant from IISER Berhampur.}

\appendix
\section[\appendixname~\thesection]{Adiabatic Evolution of TIM}\label{appA}
The energies at the end of stroke $i$ are calculated using the expresssion
\begin{equation}
{E}_i = \sum_{k} \text{Tr} [H_k^i \rho_k^i].
\end{equation}
where $i = {A},{B},{C},{D}$.
\begin{itemize}
\item[(i)] {At B}: The density matrix is given by Equation \eqref{eqn_rhoB}
and the Hamiltonian in the diagonal basis takes the form
\begin{equation}
H_{k}(h_1) = \begin{bmatrix}
 -\epsilon_k(h_1) & 0 & 0 & 0 \\ 0 & 0 & 0 & 0\\ 0 & 0 & 0 & 0\\ 0 & 0 & 0 & \epsilon_k(h_1) 
\end{bmatrix}.
\end{equation}
The energy ${E}_B$ can be calculated as follows
\begin{equation}
\text{Tr} [H_{k}(h_1) \rho_k^{B}] = \frac{ \epsilon_k(h_1)}{Z(h_1)} \left(e^{-\beta_H \epsilon_k(h_1)} - e^{\beta_H \epsilon_k(h_1)} \right)
\end{equation}
or
\begin{eqnarray}
{E}_B &=& \sum_k \text{Tr} [H_{k}(h_1) \rho_k^{B}]\\
&=& \sum_k \frac{ \epsilon_k(h_1)}{Z(h_1)} \left(e^{-\beta_H \epsilon_k(h_1)} - e^{\beta_H \epsilon_k(h_1)} \right)
\end{eqnarray}

\item[(ii)] {{At C}}: If the evolution is purely adiabatic, the populations in the eigenenergy levels do not change, resulting in
\begin{equation}
\rho_k^{C,adia} = \rho_k^B = \begin{bmatrix}
 \frac{e^{\beta_H \epsilon_k(h_1)}}{Z(h_1)} & 0 & 0 & 0 \\ 0 & \frac{1}{Z(h_1)} & 0 & 0\\ 0 & 0 & \frac{1}{Z(h_1)} & 0\\ 0 & 0 & 0 & \frac{e^{- \beta_H \epsilon_k(h_1)}}{Z(h_2)}
\end{bmatrix}
\end{equation}
and 
\begin{equation}
H_{k}(h_2) = \begin{bmatrix}
 -\epsilon_k(h_2) & 0 & 0 & 0 \\ 0 & 0 & 0 & 0\\ 0 & 0 & 0 & 0\\ 0 & 0 & 0 & \epsilon_k(h_2) 
\end{bmatrix}
\end{equation}
with $\epsilon_k(h_2) = 2\sqrt{(h_2 - \cos k)^{2} + {\sin k}^{2}}$. \\
Therefore,
\begin{equation}
{E}_C^{adia} = \sum_k \frac{ \epsilon_k(h_2)}{Z(h_1)} \left(e^{-\beta_H \epsilon_k(h_1)} - e^{\beta_H \epsilon_k(h_1)} \right)
\end{equation}

\item[(iii)] {{At D}}: This energy can be calculated similarly to that at ${B}$ so that
\begin{equation}
{E}_D = \sum_k \frac{ \epsilon_k(h_2)}{Z(h_2)} \left(e^{-\beta_C \epsilon_k(h_2)} - e^{\beta_C \epsilon_k(h_2)} \right)
\end{equation}
\item[(iv)] {{At A}}: Following the same procedure used to calculate the energy ${E}_{C}^{adia}$ in order to find the energy ${E}_{A}^{adia}$, we obtain
\begin{equation}
{E}_A^{adia} = \sum_k \frac{ \epsilon_k(h_1)}{Z(h_2)} \left(e^{-\beta_C \epsilon_k(h_2)} - e^{\beta_C \epsilon_k(h_2)} \right)
\end{equation}

\end{itemize}

The input heat energy absorbed by the WM in the non-unitary stroke ${A} \to {B}$ can be easily calculated, and is given by 
\begin{eqnarray}
\mathcal{Q}_{in}^{adia} &=& {E}_B - {E}_A^{adia} \nonumber\\
&=& \sum_k \epsilon_k(h_1) \lbrace \frac{\left(e^{-\beta_H \epsilon_k(h_1)} - e^{\beta_H \epsilon_k(h_1)} \right) }{Z(h_1)} \nonumber\\
&-& \frac{\left(e^{-\beta_C \epsilon_k(h_2)} - e^{\beta_C \epsilon_k(h_2)} \right)}{Z(h_2)} \rbrace
\end{eqnarray}

Similarly,
\begin{eqnarray}
\mathcal{Q}_{out}^{adia} &=& {E}_D - {E}_C^{adia} \nonumber\\
&=& \sum_k \epsilon_k(h_2) \lbrace\frac{\left(e^{-\beta_C \epsilon_k(h_2)} - e^{\beta_C \epsilon_k(h_2)} \right)}{Z(h_2)} \nonumber\\
&-& \frac{\left(e^{-\beta_H \epsilon_k(h_1)} - e^{\beta_H \epsilon_k(h_1)} \right)}{Z(h_1)} \rbrace .
\end{eqnarray}

We can now calculate the output work of the engine in the adiabatic limit, which is given by
\begin{eqnarray}
&\mathcal{|W|}_{adia} &= \sum_k \left( \epsilon_k(h_1)-\epsilon_k(h_2) \right) \nonumber\\
&\lbrace& \frac{\left(e^{-\beta_H \epsilon_k(h_1)} - e^{\beta_H \epsilon_k(h_1)} \right) }{Z(h_1)}  \nonumber\\
&-& \frac{\left(e^{-\beta_C \epsilon_k(h_2)} - e^{\beta_C \epsilon_k(h_2)} \right)}{Z(h_2)} \rbrace
\end{eqnarray}
resulting in
\begin{eqnarray}\label{eqn_tilde_eta}
&\eta_{adia} =  
1 - \frac{\sum_k \epsilon_k(h_2)\lbrace\frac{\left(e^{-\beta_H \epsilon_k(h_1)} - e^{\beta_H \epsilon_k(h_1)} \right) }{Z(h_1)} - \frac{\left(e^{-\beta_C \epsilon_k(h_2)} - e^{\beta_C \epsilon_k(h_2)} \right)}{Z(h_2)} \rbrace}{\sum_k \epsilon_k(h_1)\lbrace\frac{\left(e^{-\beta_H \epsilon_k(h_1)} - e^{\beta_H \epsilon_k(h_1)} \right) }{Z(h_1)} - \frac{\left(e^{-\beta_C \epsilon_k(h_2)} - e^{\beta_C \epsilon_k(h_2)} \right)}{Z(h_2)} \rbrace}.
\end{eqnarray}
\section[\appendixname~\thesection]{Power Output for BEQE}\label{appB}
\begin{figure}[h]

\includegraphics[width=9cm]{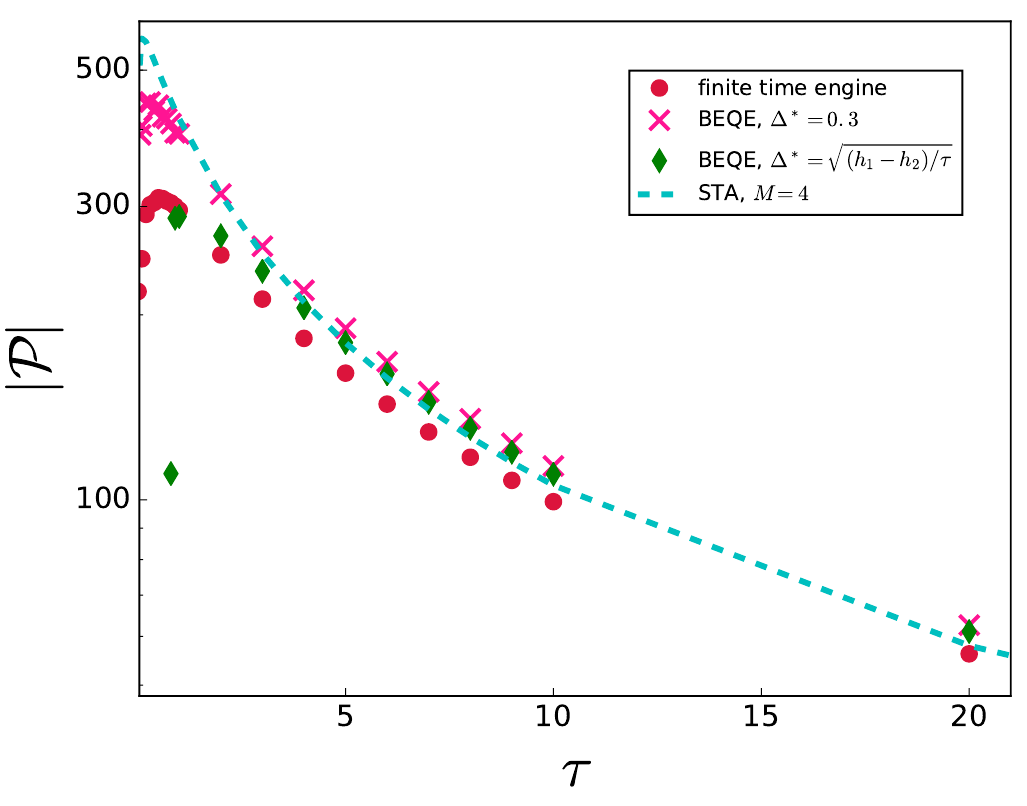}
\caption{$\mathcal{|P|}$ is plotted as a function of $\tau$ using different values of $\Delta^*$. All parameters are same as in Figure  \ref{fig_del_eta}. We have set $\tau_H + \tau_C = 4$. }
\label{fig_power}
\end{figure}

\begin{figure}[h]

\includegraphics[width=9cm]{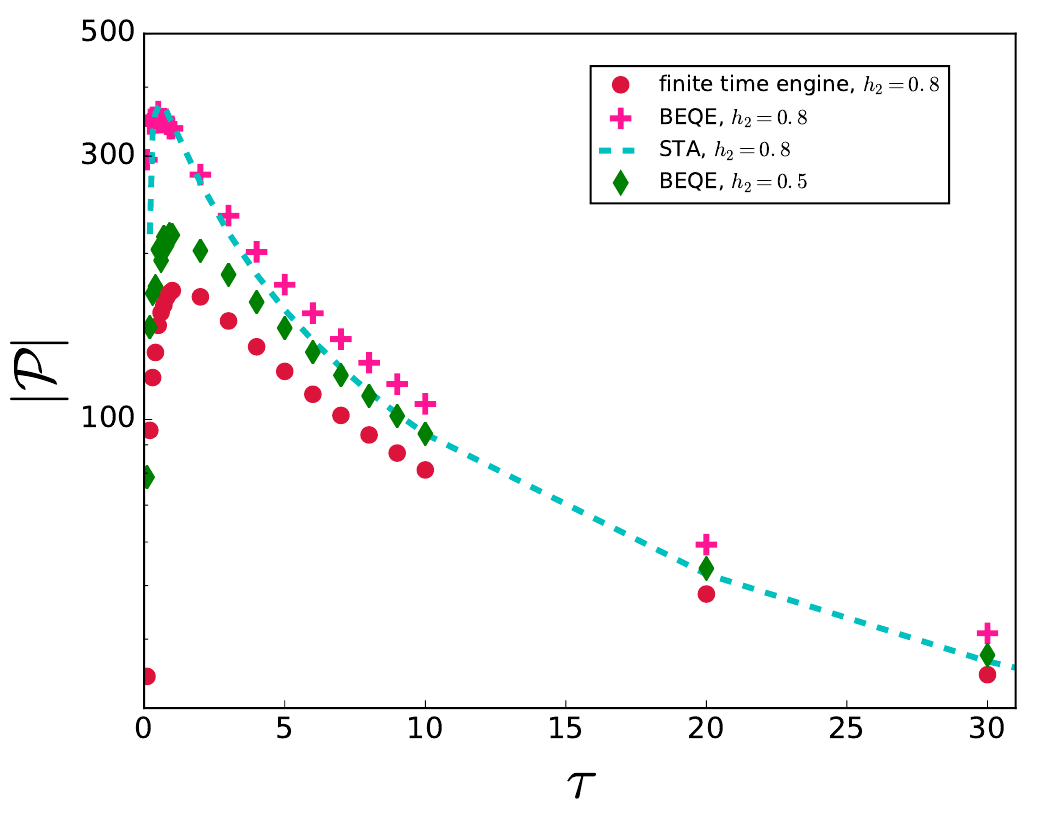}
\caption{$\mathcal{|P|}$ is plotted as a function of $\tau$ for $h_2 < 1$. Here, $L = 1000, h_1 = 10, T_H = 20, T_C = 1, \tau_1 = \tau_2, \Delta^*(h_2 = 0.5) = 1.02, \Delta^*(h_2 = 0.8) = 0.48$. We have set $\tau_H + \tau_C = 4$. }
\label{fig_powernc}
\end{figure}

{{
The power output for an engine is defined as
\begin{equation}
\mathcal{P} = \frac{- \mathcal{W}}{\tau_{\text{total}} }
\end{equation}
where $\tau_{\text{total}} = \tau_1 + \tau_2 + \tau_H + \tau_C$, with $\tau_H (\tau_C)$ 
being the time duration of
the {A} $\rightarrow$ {B} ({C} $\rightarrow$ {D}) non-unitary stroke. As shown in Figures \ref{fig_power} and \ref{fig_powernc}, the BEQE
outperforms finite-time engines without controls and with STA for a
wide range of $\tau$ values.}}

\section[\appendixname~\thesection]{BEQE Using LTIM WM}\label{appC}
With LTIM as the WM of the quantum Otto cycle, the transverse field $B_x$ is changed from an $h_1$ to an $h_2$ value during the unitary strokes. Following the technique presented here, bath engineering can be implemented by evaluating the corresponding $\Delta^*$ using the Kibble--Zurek mechanism and then choosing an appropriate cut-off for bath spectral functions such that  energy gaps which are less than $\Delta^*$ do not participate in the dynamics. LTIM falls under the same universality class as that of TIM.

Let us discuss the implementation of single-stroke bath engineering in LTIM.
When performing bath engineering, some levels will not be allowed to thermalize, depending upon the energy gap.
There are $2^L$ energy levels for a system size $L$ and thus $2^{L}-1$ energy gaps. Those energy levels having gaps less than the threshold value of $\Delta^* = \left( \frac{\nu z (h_1 - h_2)}{\tau}\right)^{\frac{\nu z}{1 + \nu z}}$ will not thermalize. 
On the other hand, those with energy gaps greater than $\left( \frac{\nu z (h_1 - h_2)}{\tau}\right)^{\frac{\nu z}{1 + \nu z}}$ will thermalize according to the equation
\begin{equation}
p_{i} = p_{i-1} e^{-(E_{i}-E_{i-1})/T}
\end{equation}
where $p_{i}$ and $p_{i-1}$ are the populations in the $i^{th}$ and $(i-1)^{th}$ energy levels.

\begin{itemize}
\item In the {C $\rightarrow$ D} non-unitary stroke, the Hamiltonian is with a transverse field $h_2$. 
  The energy gaps are compared with the $\Delta^*$ value. Those energy levels having gaps greater than $\Delta^*$ are allowed to interact with the decaying bath in the {C $\rightarrow$ D} stroke and thus thermalize according to 

\begin{equation}
\frac{p_{i+1}^D}{p_{i}^D} = e^{-(E_{i+1} - E_{i})/T_c}
\end{equation}
where $E_{i+1} - E_{i} > \Delta^*$.

In order to apply bath engineering in the case of gaps that are less than $\Delta^*$, i.e, when $E_{i+1} - E_{i} < \Delta^*$, we have two possibilities.
\begin{itemize}
\item[(i)] If $E_{i} - E_{i-1} > \Delta^*$, the populations are determined by the condition 
\begin{equation}\label{con}
p_{i}^{D} + p_{i-1}^{D} = p_{i}^C + p_{i-1}^C.
\end{equation}
\item[(ii)] If $E_{i} - E_{i-1} < \Delta^*$, the $i^{th}$ level does not interact with any other level, leading to $p_i^D = p_i^C$.
\end{itemize}
Solving these system of equations, along with the condition that $\sum_i p_i^D = 1$, we obtain the populations of all the other energy levels at {D}. Thus, we have $\rho^{\prime D}$, which is the state reached after carrying out bath engineering.

\item From {D} to {A}, $h_2$ is changed back to $h_1$ from $\rho^{\prime D}$ using the evolution equation 
\begin{equation}
\frac{d\rho}{dt} = -i [H, \rho]
\end{equation}
 which gives the new density matrix at {A}, $\rho^{\prime A}$.

\item At {A}, the system with the Hamiltonian $H(h_1)$ is connected to the energizing bath. All energy levels interact with each other, resulting in the steady state at {B}. 

\item {B $\rightarrow$ C} stroke, $h_1$ changed to $h_2$ by evolving the system from $\rho^B$ to obtain $\rho^C$.
\end{itemize}

\begin{figure}[h]

\includegraphics[width=9cm]{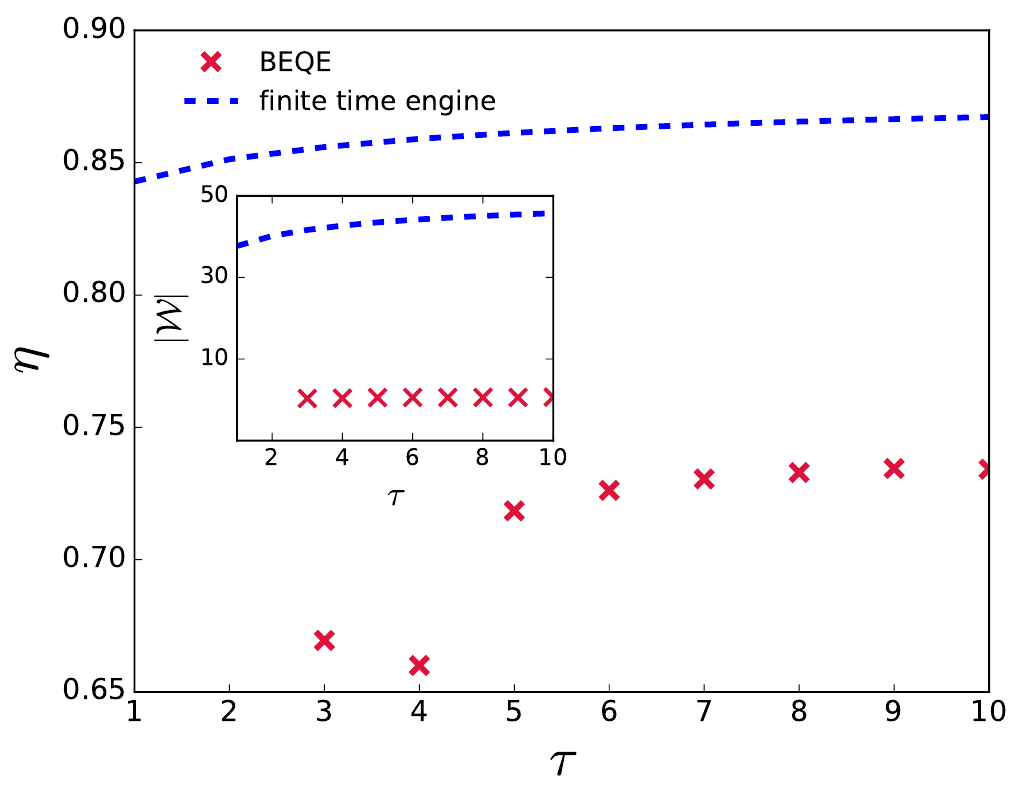}
\caption{$\eta$ is plotted as a function of $\tau$ for the BEQE using LTIM as the WM compared with a finite-time engine. Inset: $\mathcal{|W|}$ as a function of $\tau$. The parameters used are $L = 6, h_1 = 10, h_2 = 0.75, T_H = 500, T_C = 0.1, \tau_1 = \tau_2 = \tau$.}
\label{fig_ltim_eta}
\end{figure}

Now, for the bath-engineered engine,
\begin{eqnarray}
\mathcal{Q}_{in}^{\prime} &=& {E}_B - {E}_{A}^{\prime}\\
\mathcal{Q}_{out}^{\prime} &=& {E}_D^{\prime} - {E}_{C}\\
\mathcal{W}^{\prime} &=& - (\mathcal{Q}_{in}^{\prime} + \mathcal{Q}_{out}^{\prime})\\
\eta^{\prime} &=& -\mathcal{W}^{\prime}/\mathcal{Q}_{in}^{\prime}
\end{eqnarray}
Numerical analysis suggests that bath engineering failed to improve the performance of the engine in this case (see Figure \ref{fig_ltim_eta}). 
We also investigated the implementation of bath engineering in
both energizing and decaying strokes, and again failed to improve the
performance of the engine.

\newpage
\end{document}